\documentclass[8.5pt,twoside,twocolumn]{article}
\usepackage[super,sort&compress,comma]{natbib} 
\usepackage[version=3]{mhchem}
\usepackage{balance}
\usepackage{times,mathptm}
\usepackage{sectsty}
\usepackage{graphicx} 
\usepackage{lastpage}
\usepackage[format=plain,justification=raggedright,singlelinecheck=false,font=small,labelfont=bf,labelsep=space]{caption} 
\usepackage{fancyhdr}
\begin{document}

\thispagestyle{plain}
\fancypagestyle{plain}{
\fancyhead[L]{\includegraphics[height=8pt]{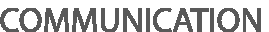} }
\fancyhead[C]{\hspace{-1cm}\includegraphics[height=20pt]{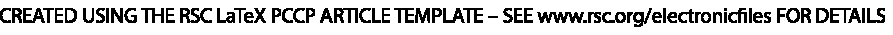}}
%\fancyhead[R]{\hspace{10cm}\vspace{-0.25cm}\includegraphics[height=10pt]{headers/RH}}
\renewcommand{\headrulewidth}{1pt}}
\renewcommand{\thefootnote}{\fnsymbol{footnote}}
\renewcommand\footnoterule{\vspace*{1pt}% 
\hrule width 3.4in height 0.4pt \vspace*{5pt}} 
\setcounter{secnumdepth}{5}

\makeatletter 
\renewcommand\@biblabel[1]{#1}      
\renewcommand\@makefntext[1]% 
{\noindent\makebox[0pt][r]{\@thefnmark\,}#1}
\makeatother 
\renewcommand{\figurename}{\small{Fig.}~}
\sectionfont{\large}
\subsectionfont{\normalsize} 

\fancyfoot{}
\fancyfoot[LO,RE]{\vspace{-7pt}\includegraphics[height=9pt]{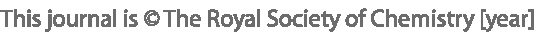}}
%\fancyfoot[CO]{\vspace{-7.2pt}\hspace{12.2cm}\includegraphics{headers/RF}\\[-13pt]{\scriptsize{\textsf{Typeset:\today}}}}
%\fancyfoot[CE]{\vspace{-7.5pt}\hspace{-13.5cm}\includegraphics{headers/RF}\\[-13pt]{\scriptsize{\textsf{Typeset:\today}}}}
\fancyfoot[CO]{\vspace{-7.2pt}\hspace{12.2cm}\\[-5pt]{\scriptsize{\textsf{Typeset:\today}}}}
\fancyfoot[CE]{\vspace{-7.5pt}\hspace{-13.5cm}\\[-5pt]{\scriptsize{\textsf{Typeset:\today}}}}
\fancyfoot[RO]{\footnotesize{\sffamily{1--\pageref{LastPage} ~\textbar \hspace{2pt}\thepage}}}
%\fancyfoot[LE]{\footnotesize{\sffamily{\thepage~\textbar\hspace{3.45cm} 1--\pageref{LastPage}}}}
\fancyfoot[LE]{\footnotesize{\sffamily{\thepage~\textbar 1--\pageref{LastPage}}}}
\fancyhead{}
\renewcommand{\headrulewidth}{1pt} 
\renewcommand{\footrulewidth}{1pt}
\setlength{\arrayrulewidth}{1pt}
\setlength{\columnsep}{6.5mm}
\setlength\bibsep{1pt}

\twocolumn[
 \begin{@twocolumnfalse}
\noindent\LARGE{\textbf{Defect-dependent colossal negative thermal expansion in UiO-66(Hf) metal--organic framework$^\dag$}}
\vspace{0.6cm}

\noindent\large{\textbf{Matthew J.\ Cliffe,\textit{$^{a}$} Joshua A.\ Hill,\textit{$^{a}$} Claire A.\ Murray,\textit{$^{b}$} Fran\c{c}ois-Xavier Coudert,\textit{$^{c}$} and\\Andrew L.\ Goodwin$^{\ast}$\textit{$^{a}$} }}\vspace{0.5cm}

\noindent\textit{\small{\textbf{Received Xth XXXXXXXXXX 20XX, Accepted Xth XXXXXXXXX 20XX\newline
First published on the web Xth XXXXXXXXXX 200X}}}

\noindent \textbf{\small{DOI: 10.1039/b000000x}}
 \end{@twocolumnfalse} \vspace{0.6cm}

 ]

\noindent\textbf{Thermally--densified hafnium terephthalate UiO-66(Hf) is shown to exhibit the strongest isotropic negative thermal expansion (NTE) effect yet reported for a metal--organic framework (MOF). Incorporation of correlated vacancy defects within the framework affects both the extent of thermal densification and the magnitude of NTE observed in the densified product. We thus demonstrate that defect inclusion can be used to tune systematically the physical behaviour of a MOF.}
\section*{}
\vspace{-1cm}
\footnotetext{\dag~Electronic Supplementary Information (ESI) available: Synthesis, experimental methods, and sample characterisation; X-ray powder diffraction refinement details.}
\footnotetext{\textit{$^{a}$~Department of Chemistry, University of Oxford, Inorganic Chemistry Laboratory, South Parks Road, Oxford OX1 3QR, U.K. Fax: +44 1865 274690; Tel: +44 1865 272137; E-mail: andrew.goodwin@chem.ox.ac.uk}}
\footnotetext{\textit{$^{b}$~Diamond Light Source, Ltd., Harwell Science and Innovation Campus, Didcot,
Oxfordshire, OX11 0DE, UK. }}
\footnotetext{\textit{$^{c}$~PSL Research University, Chimie ParisTech--CNRS, Institut de Recherche de Chimie Paris, 75005 Paris, France. }}
\section*{Introduction}

It has long been recognised that defects and their correlations play a central role in determining the physical properties of many functional materials: the mass transport pathways of fast-ion conductors,\cite{Tealdi2010,Ruiz-Morales2006} charge localisation in high-temperature superconductors,\cite{Bozin1999,Bobroff2002} and emergence of polar nanoregions in relaxor ferroelectrics\cite{Pasciak2012} are just three examples. The recent discovery that correlated defects can be systematically introduced into the structures of some canonical metal--organic frameworks (MOFs)\cite{Cliffe2014} suggests the possibility of establishing defect/property relationships in this broad family of materials long-favoured for its chemical versatility.\cite{Horike2009,Banerjee2008,Cairns2013} Perhaps the most obvious benefit of determining such relationships would be the ability to exploit defects and non-stoichiometry to develop MOFs with particularly attractive or otherwise-inaccessible physical properties.

The few recent studies of defective MOFs have focussed either on the nature of defect inclusion itself or on how defects influence chemical properties such as catalytic activity and guest sorption.\cite{Park2012,Vermoortele2013,Barin2014,Fang2014} Here we ask a different question: namely, is it possible that defects can influence the \emph{physical} properties of MOFs? In attempting to answer this question, we focus in this proof-of-principle study on the thermomechanical behaviour of the canonical hafnium terephthalate MOF UiO-66(Hf).\cite{Jakobsen2012} Our choice of this specific pairing of physical property and chemical system is motivated by the following considerations. First, the response of framework geometry to changes in temperature is at once both straightforwardly measured and strongly characteristic of the underlying elastic behaviour of the material in question.\cite{Ogborn2012,Collings2014} Second, the empirical propensity for MOFs to exhibit anomalous mechanical effects such as thermal contraction (\emph{i.e.}\ negative thermal expansion, NTE; Refs.~\citenum{Zhou2008,Wu2008a,Wu2014}) has led to a general expectation that such properties will be especially sensitive to the existence and nature of structural defects.\cite{Cairns2013} And third, UiO-66(Hf) is the MOF for which there is arguably the greatest and best-understood chemical control over defect incorporation.\cite{Wu2013a,Vermoortele2013,Katz2013,Cliffe2014}

\begin{figure}
\centering
 \includegraphics{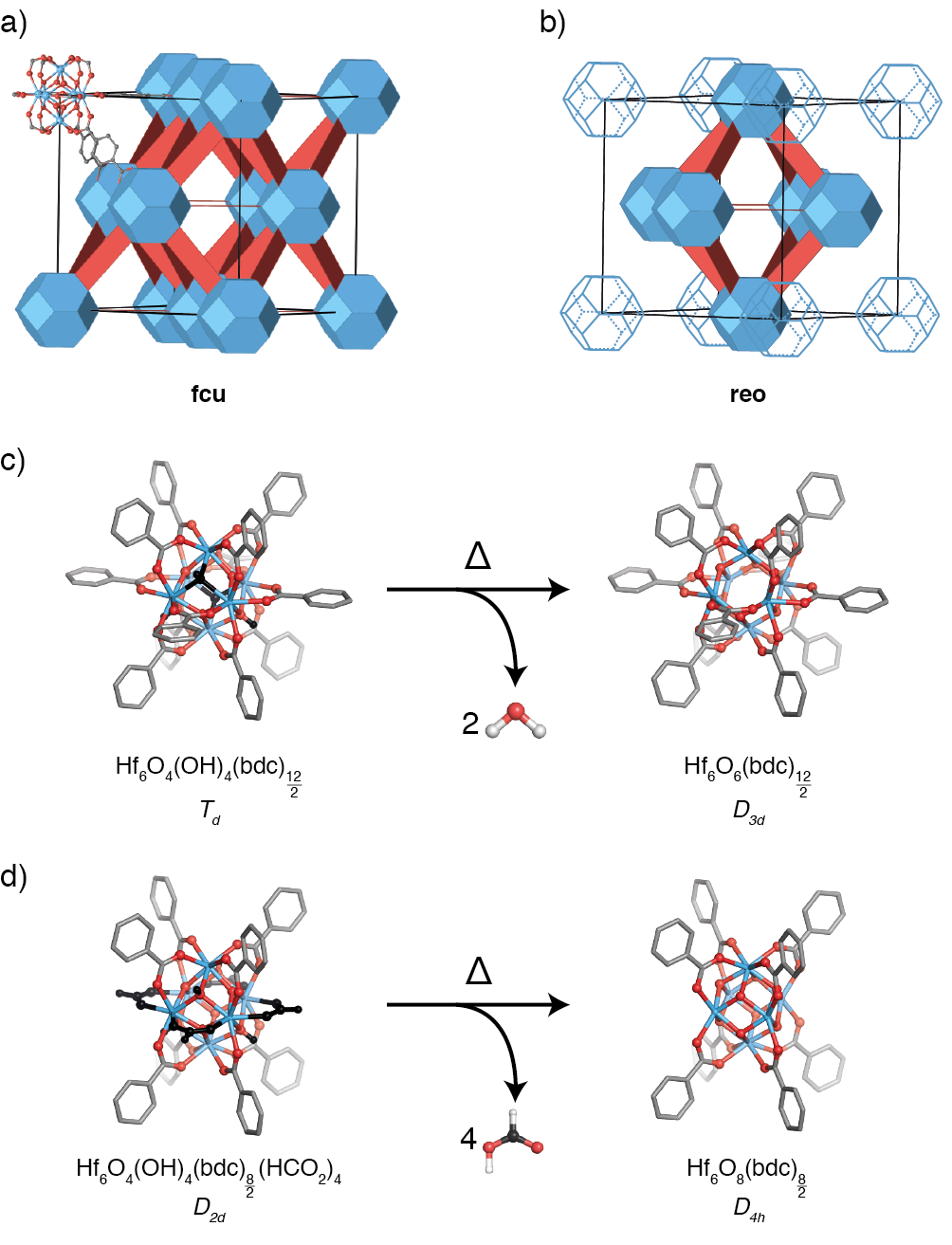}
 \caption{The ideal non-defective UiO-66(Hf) consists of oxyhydroxyhafnium(IV) clusters connected via terephthalate linkers into a face-centred cubic \textbf{fcu} topology network. A unit cell is shown in panel (a), with the Hf clusters represented as blue truncated octahedra and linkers as red rectangles. The atomic-scale structure of a single cluster is shown in detail in the top-left of the panel. (b) at high concentrations of formic acid promotes the formation of nanodomains of the primitive \textbf{reo} topology, which contains correlated linker and hafnium cluster absences. Absent hafnium clusters are shown in skeletal form. In spite of the differences in framework topology, the cluster coordination environments of \textbf{fcu} (c) and \textbf{reo} (d) UiO-66(Hf) are very similar.\cite{Cliffe2014,Wiersum2011} After ligand elimination (right-hand panels of (c),(d)) the two cluster types are distinct in both coordination and symmetry. Atoms are coloured as follows: Hf, blue; O, red; and C, grey. Atoms lost during thermal elimination are coloured black.
 \label{fig1}}
 \end{figure}

The idealised structure of non-defective UiO-66(Hf) consists of a face-centred cubic array of oxyhydroxyhafnium(IV) clusters connected via terephthalate (benzene-1,4-dicarboxylate, bdc$^{2-}$) linkers [Fig.~\ref{fig1}(a)].\cite{Jakobsen2012} The particular defects we study here are introduced by substituting formate for terephthalate during framework synthesis. As a monotopic (capping) ligand, formate reduces the connectivity of the framework. This, in turn, encourages vacancies of entire hafnium clusters. Formate content can be increased at least as far as the situation where, of the 12 carboxylate ligands of each hafnium cluster, four ligands are formates and eight are terephthalate carboxylates [Fig.~\ref{fig1}(b)].\cite{Vermoortele2013,Cliffe2014} While this substitution results in substantive structural modification over the unit-cell (nanometre) length scale---indeed one in four hafnium clusters are vacant in this case---the actual coordination environment within the clusters is essentially indistinguishable in the defective and defect-free modifications.\cite{Oien2014,Cliffe2014,Wiersum2011} In order to amplify the structural consequences of defect incorporation we exploit the high-temperature ligand elimination reaction known to occur in this system.\cite{Vermoortele2013,Shearer2013,Valenzano2011} Whereas this step involves elimination of H$_2$O for defect-free UiO-66, the lability of non-bridging formate ligands is such that it is formic acid that is eliminated in our defective UiO-66, resulting in markedly different cluster geometries for the two cases [Fig.~\ref{fig1}(c,d)].\cite{Valenzano2011,Vandichel2014}

Making use of \emph{in situ} variable-temperature synchrotron X-ray diffraction measurements of a series of UiO-66(Hf) samples with a range of defect concentrations, we proceed to demonstrate the following. First, ligand elimination results in framework densification, the magnitude of which is dependent on defect concentration. Quantum mechanical calculations help interpret the microscopic origin of this volume reduction. Second, all densified-UiO-66(Hf) samples exhibit isotropic NTE behaviour that is many times stronger than that of any other MOF, and in one case even satisfies the condition of ``colossal'' NTE (its volumetric coefficient of thermal expansion $\alpha_V={\rm d}V/V{\rm d}T\simeq-100$\,MK$^{-1}$). This is the first report of colossal NTE in an isotropic MOF. Third, the magnitude of NTE can also be tuned systematically by defect concentration. Our paper concludes with a discussion of the likely mechanisms by which defect inclusion might influence elastic properties of this particular MOF, the challenges posed for understanding NTE in defective materials, and the implications for exploiting defect/property relationships in other MOF systems.

% This framework has the nominal composition Hf$_6$O$_4$(OH)$_4$(bdc)$_6$ (bdc$^{2-}$ = 1,4-benzenedicarboxylate or terephthalate).

\section*{Results and discussion}

\begin{figure}
\centering
 \includegraphics{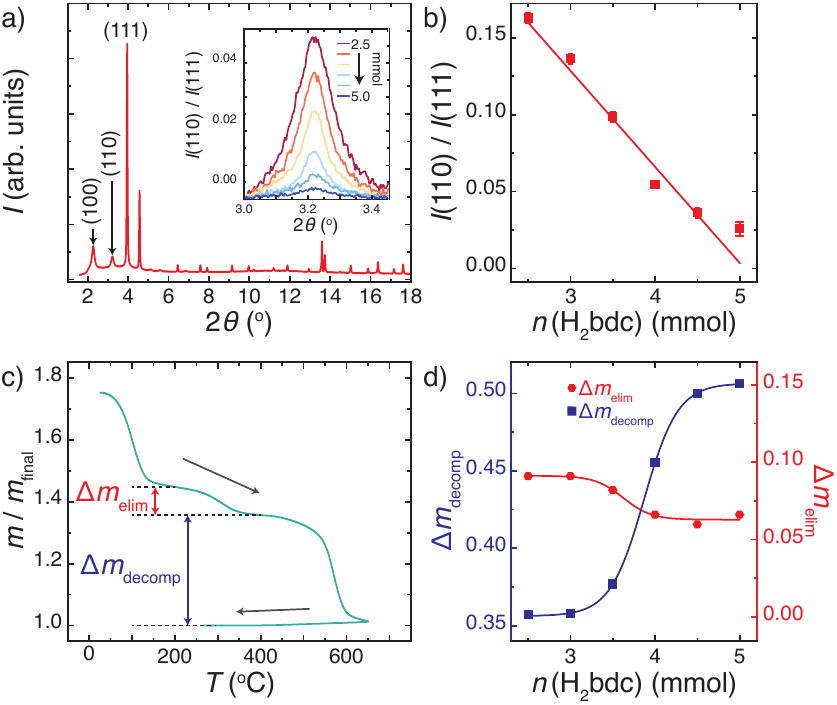}
 \caption{
(a) A diffraction pattern of the most defective sample of UiO-66(Hf). Structure diffuse scattering located at primitive superlattice positions is a signature of the presence of correlated defects in UiO-66(Hf). The most intense of these broad reflections are the (100) and (110). On increasing the quantity of H$_2$bdc in the reaction mixture, the concentrations of defects decreases, and so the relative intensity of these primitive reflections decreases ((a) inset and (b)). (c) The TGA trace of UiO-66(Hf) shows three steps corresponding to, in turn, guest volatilisation, ligand elimination and framework decomposition. (d) The decrease in defect concentration is evident in the variation with bdc concentration of normalised mass losses both on ligand elimination ($\mathrm{\Delta}m_\textrm{elim}$), which decreases, and framework decomposition ($\mathrm{\Delta}m_\textrm{decamp}$). Fitted curves are included as guides to the eye. \label{fig2}}
\end{figure}

Having used the method of Ref.~\citenum{Cliffe2014} to prepare a series of six UiO-66(Hf) samples with a range of defect concentrations, we sought first to establish that our samples included defects of the type discussed above. X-ray powder diffraction measurements revealed the existence of structured diffuse scattering in the form of broad superlattice reflections that are symmetry-forbidden in the defect-free system but are associated with nanodomains of correlated cluster vacancies in defective UiO-66 [Fig.~\ref{fig2}(a)]. The ratio of the intensity of this diffuse scattering to that of the parent reflections is a measurable quantity that is sensitive to defect concentration. We find an approximately linear relationship between this ratio and the quantity of terephthalic acid used during synthesis, suggesting that our ensemble of defective UiO-66(Hf) samples spans a suitable range of defect concentrations [Fig.~\ref{fig2}(a,b)].

The onset temperature of ligand elimination was determined using thermogravimetric analysis (TGA). The TGA traces of all our samples are qualitatively similar and can be interpreted on the basis of previous studies of this system:\cite{Valenzano2011,Vermoortele2013,Shearer2014} each reveals three primary stages of mass loss, corresponding in turn to adsorbate volatilisation ($\sim$100\,$^\circ$C), ligand elimination (250--350\,$^\circ$C) and framework decomposition (550--600\,$^\circ$C) [Figs.~\ref{fig2}(c), ESI Figs.~1\&2\dag]. Qualitative trends in defect concentrations can be inferred from the relative mass losses associated with the second and third stages [$\Delta m_{\rm elim}$ and $\Delta m_{\rm decomp}$, respectively, in Fig.~\ref{fig2}(c,d)]. Framework decomposition corresponds to the loss of terephthalate (leaving behind HfO$_2$) and so larger values of $\Delta m_{\rm decomp}$ are expected for the least defective examples, which should contain the highest relative concentrations of terephthalate. This is precisely what we observe [Fig.~\ref{fig2}(d)]. Likewise the value of $\Delta m_{\rm elim}$ should decrease as the defect concentration is reduced (since H$_2$O is eliminated instead of the heavier HCOOH), and again this trend is observed in practice. So both X-ray diffraction and TGA measurements point to a systematic variation in defect concentration across our series of UiO-66(Hf) samples.

In order to determine the effect of defect concentration on thermomechanical behaviour, we measured variable-temperature X-ray powder diffraction patterns for all six samples. Because we were interested in determining the thermal expansivity of the phases produced by ligand elimination, we paid particular care to ensuring that our samples were heated sufficiently to drive this elimination step but not so far as to initiate framework decomposition. In practice this dictated an experimental protocol of heating each sample to a temperature of 340\,$^\circ$C and then measuring thermal expansivity on subsequent cooling to 100\,$^\circ$C. So the temperature range probed in these experiments corresponds to the first two stages observed in the TGA measurements described above.

\begin{figure}[b]
\centering
 \includegraphics{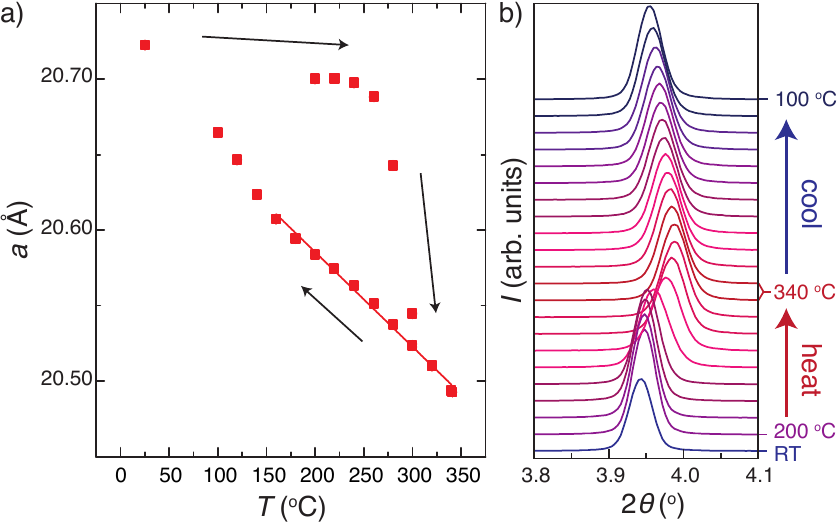}
 \caption{(a) Evolution of the lattice parameter of the UiO-66(Hf) sample synthesised with the largest quantity of H$_2$bdc (5.0\,mmol) --- i.e. the lowest defect concentration --- on heating from room temperature to 340\,$^\circ$C then cooling to 100\,$^\circ$C. A linear fit to the lattice parameter over the temperature range 340--160\,$^\circ$C is shown, for the determination of $\alpha_V$. (b) The evolution of the most intense peak, the (111) reflection, throughout this process.\label{fig3}}
\end{figure}

For all samples the same qualitative behaviour was observed (although, as will form the focus of subsequent discussion, the magnitudes of these changes varied from sample to sample). On heating through the first stage of mass loss (\emph{i.e.}\ up to 250\,$^\circ$C) the most noticeable changes to the diffraction pattern were in the relative intensities of reflections sensitive to solvent occupation within the framework pores. There was only a very small change ($\sim$0.1\%) in the size of the unit cell within this regime. These observations are consistent with our interpretation of the TGA trace and with previous investigations.\cite{Valenzano2011,Vermoortele2013} In contrast, the ligand elimination process that occurs between 250 and 350\,$^\circ$C resulted in an order-of-magnitude larger decrease in the cubic unit cell parameter that has not previously been observed in defect-free samples [Fig.~\ref{fig3}(a)]. The diffraction pattern of the densified framework formed at this higher temperature can still be accounted for by a model based on the ambient-temperature structure, adjusted for the change in lattice dimensions; however the scattering intensity in high-angle Bragg peaks was much reduced such that detailed structural refinement did not prove possible. On subsequent cooling from 340 to 100\,$^\circ$C all samples showed NTE [Fig.~\ref{fig3}(a,b)]. The two processes of densification and NTE, together with their dependence on defect concentration, are discussed in turn below.

\subsection*{Thermal densification}

In the absence of a diffraction-based structural model for the densified framework, we turned to quantum mechanical calculations to verify that the structural changes associated with formate elimination were consistent with the magnitude of volume collapse observed experimentally. From a computational viewpoint it would have been unfeasible to use \emph{ab initio} methods to relax large atomistic configurations describing the nanodomain defect structure reported in Ref.~\citenum{Cliffe2014}. Instead we focussed on an ordered defect structure corresponding to substitution of four equatorial carboxylate sites by formate on each hafnium cluster; this is the `ordered-{\bf reo}' model of Ref.~\citenum{Cliffe2014}. For this model, ligand elimination corresponds to the chemical process
\begin{center}
Hf$_6$O$_4$(OH)$_4$(bdc)$_4$(OOCH)$_4\rightarrow$ Hf$_{6}$O$_8$(bdc)$_4$ + 4HCOOH$\uparrow$.
\end{center}
\emph{In silico} relaxation of structural models for the framework compositions represented on either side of this reaction equation suggested that a 1.0\% change in lattice parameters is expected upon HCOOH elimination: $a=20.862$\,\AA\ for Hf$_6$O$_4$(OH)$_4$(bdc)$_4$(OOCH)$_4$ and 20.652\,\AA\ for Hf$_6$O$_8$(bdc)$_4$.\footnote[3]{The tendency for calculations to overestimate lattice parameters for organic and organic-containing materials is well-known phenomenon and arises due to the difficulty of correctly accounting for dispersive interactions. It does not impact the analysis of structural and lattice parameter trends in families of materials of similar composition and porosity.\cite{Haigis2014}} In contrast, it is already known that elimination of water from defect-free UiO-66 does not result in any significant volume change.\cite{Vandichel2014} By inspection of the relaxed structural model for Hf$_6$O$_8$(bdc)$_4$, we conclude that distortions of the Hf$_6$ octahedra are responsible for this framework densification. Elimination of HCOOH amplifies the tetragonal distortion of Hf$_6$ octahedra, reducing the effective octahedral volume by 4.2\% and bending the cluster---bdc---cluster connection so as to couple local distortion to macroscopic volume contraction [ESI Fig.~4 \dag]. The relaxed structural models are available as crystallographic information files as part of the ESI.\dag

\begin{figure}
\centering
 \includegraphics{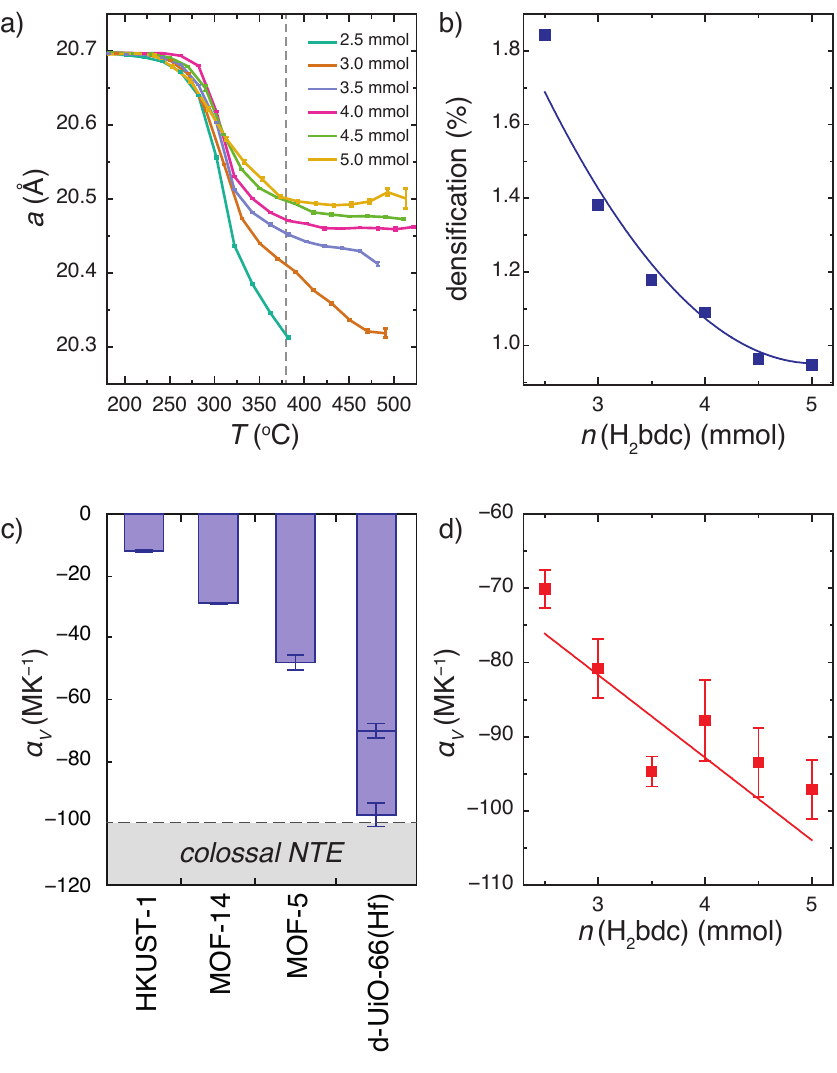}
 \caption{(a) The evolution of the lattice parameter through the ligand elimination process for a series of samples with different defect concentrations. To correct for minor variations in the onset temperature of ligand elimination between sample, the curves are offset in temperature such that the maximum change in volume corresponds to 300\,$^\circ$C. (b) The densification (percentage linear decrease) after ligand elimination, measured at 380\,$^\circ$C, highlighted in (a) by a dashed grey line. (c) Coefficients of thermal expansion for isotropic NTE MOFs. The two values given for densified-UiO-66(Hf) correspond to upper and lower defect concentrations.\cite{Zhou2008,Wu2008a,Wu2014} (d) Variation in NTE behaviour as a function of decreasing defect concentration (increasing quantity of H$_2$bdc in the reaction mixture). \label{fig4}}\end{figure}

If, as this analysis would suggest, the process of framework densification depends on the presence of structural defects, then we should expect stronger densification in UiO-66(Hf) samples with larger defect concentrations. By heating our samples to sufficiently-high temperatures to ensure complete densification (and working within the constraints of available synchrotron beam time), we found this expected property/defect relationship to be borne out in practice [Fig.~\ref{fig4}(a,b)]. Our most defective samples showed densifications that were larger by $\sim$1\% (of the total lattice parameter) than for the least defective samples; this result is consistent not only with the results of our \emph{ab initio} calculations but also with previous experimental reports.\cite{Vandichel2014} 

\subsection*{Negative thermal expansion}

Perhaps the most intriguing physical property of densified-UiO-66(Hf) is its NTE response, which is evident in the increase of the cubic unit cell parameter on cooling from 340\,$^\circ$C [Fig.~\ref{fig3}(a)] and in the anomalous thermal shifts of Bragg peaks in the raw diffraction data themselves [Fig.~\ref{fig3}(b)]. While NTE is increasingly frequently observed amongst MOFs, there are two particular aspects of the behaviour we observe here that are especially attractive: on the one hand, the cubic crystal symmetry of defective UiO-66 means that its NTE is isotropic---\emph{i.e.}, it occurs with constant magnitude in all crystal directions; and, on the other hand, the magnitude itself is really very large. Thermal expansion effects in different materials can be compared via the lattice expansivities
\begin{equation}
\alpha_V=\frac{1}{V}\left(\frac{\partial V}{\partial T}\right)_p,
\end{equation}
with typical (positive) thermal expansion of engineering materials corresponding to values of $\alpha_V$ in the range 10--30\,MK$^{-1}$.\cite{Krishnan1979} Here, the lattice parameter data of Fig.~\ref{fig3}(a) correspond to a thermal expansivity of $\alpha_V=-97(4)$\,MK$^{-1}$ over the temperature range 160--340\,$^\circ$C, which is comparable in magnitude to that of ``colossal'' thermal expansion materials ($|\alpha|\simeq100$\,MK$^{-1}$) such as Ag$_3$[Co(CN)$_6$].\cite{Goodwin2008} Although we observe an even more rapid increase in lattice parameter on further cooling from 160--100\,$^\circ$C, we cannot rule out the possibility that this feature may arise from re-uptake of guest molecules at these lower temperatures, and hence do not attach particular weight to that observation here. We have also carried out separate variable temperature diffraction measurements confirming the reversibility of the NTE behaviour [ESI Fig.~4].

Isotropic NTE behaviour of this magnitude is rare indeed. First discovered in ZrW$_2$O$_8$ in the late 1960s,\cite{Martinek1968} isotropic NTE is now known to occur in a handful of materials, such as ZrW$_2$O$_8$ itself and related substitutional variants,\cite{Martinek1968,Korthuis1995,Tallentire2013} the perovskite analogues ScF$_3$ and ReO$_3$,\cite{Rodriguez2009,Chatterji2008,Greve2010} the cuprite--structured oxides Ag$_2$O and Cu$_2$O,\cite{Tiano2003} the Zn$_x$Cd$_{1-x}$(CN)$_2$ family and anhydrous Prussian Blue analogues,\cite{Goodwin2005a,Goodwin2005,Chapman2006,Phillips2008} and the three MOFs HKUST-1, MOF-5, and MOF-14.\cite{Zhou2008,Wu2008a,Wu2014} Recent theoretical calculations suggest that isotropic NTE may occur in a wider range of MOF materials that currently experimentally reported and re-emphasise the important role of topology.\cite{BouesselduBourg2014} Of these various systems, single-network Cd(CN)$_2$ is the only colossal NTE material ($\alpha_V=-100.5(15)$\,MK$^{-1}$).\cite{Phillips2008} Despite decades of theoretical and experimental studies there is only general consensus that low-energy transverse vibrational modes play at least some role in the NTE mechanisms for most of these materials. The details of NTE mechanisms remain controversial, with new studies frequently demanding reinterpretation of long-held assumptions as and when they appear.\cite{Zhou2008,Lock2010,Lock2012,Rimmer2014} Here we avoid speculating on a detailed NTE mechanism and focus instead on the measurement and defect dependence of the phenomenon. We note simply that, with its open framework structure of high-nuclearity clusters connected by long, light, and flexible organic linkers, densified-UiO-66(Hf) certainly shares many of the basic design features thought to favour low-energy NTE modes. Fig.~\ref{fig4}(d) places the isotropic NTE behaviour of UiO-66(Hf) in the context of the three other MOFs known to show the effect.\cite{Zhou2008,Wu2008a,Wu2014}

As anomalous as the thermal expansion behaviour of densified-UiO-66(Hf) is, our key interest is in the possibility that the magnitude of NTE might be sensitive to defect concentration. What sensitivity, if any, do we actually expect? An earlier study of the geometric contribution to NTE in molecular frameworks suggested that defects of the type thought to occur in UiO-66(Hf) should actually encourage NTE because they act to increase the relative contribution of NTE vibrational modes to the overall framework dynamics.\cite{Goodwin2006} Likewise, even simple Maxwellian counting suggests that the balance of degrees of freedom and geometric constraints would be tipped increasingly in favour of the former (and hence NTE) as linkers are removed from a three-dimensional network.\cite{Thorpe2009} Not taken into account in either analysis, however, is the observation we make here that defects are also linked to framework densification. Empirically, one finds that higher densities usually result in weaker NTE---this is exactly what is observed in zeolites, Prussian blues, and the Zn$_x$Cd$_{1-x}$(CN)$_2$ family alike.\cite{Lightfoot2001,Goodwin2005a,Chapman2006} To complicate further this balance of competing effects, we remark that the ligand elimination process alters the point symmetry of the Hf$_6$ clusters in a way that may itself affect the dynamics---and hence the NTE---of defective UiO-66(Hf). The point symmetries of both defect-free and defective Hf$_6$ clusters ($D_{3d}$ and $D_{4h}$, respectively) are incompatible with one another, as they are with the point symmetry of the crystallographic site on which they both sit ($O_h$) [Fig.~\ref{fig1}(c,d)]. Consequently, the effective energy potential governing transverse vibrational motion---which, by analogy to better-understood NTE materials, is likely to be implicated in NTE---will not be constrained by symmetry and hence will vary from site to site throughout the defective lattice. Despite the differences in local symmetry between the \textbf{reo} nanodomains and \textbf{fcu} matrix, both phases retain their cubic symmetry on the unit cell level. The thermal expansion within each domain will therefore remain isotropic. Differences in thermal expansion between phases will lead to the development of thermal strains at the domain walls, which would in principle manifest themselves as differential peak broadening between primitive and face-centered reflections on heating. The large strains produced by the densification precluded our measurement of this comparatively subtle effect. The effect of defect concentration on the average effective potential, its asymmetry, and its variance, collectively poses an interesting challenge for computational and theoretical methods of understanding NTE lattice dynamics in this system.

What we observe in practice is that our UiO-66(Hf) samples with higher defect concentrations actually show reduced NTE effects relative to defect-poor samples. The coefficients of thermal expansion determined from our variable-temperature X-ray powder diffraction measurements range systematically from $-70(2)$ to $-97(4)$\,MK$^{-1}$ as defect concentrations are reduced [Fig.~\ref{fig4}(c)]. So any effect of increased flexibility arising from the reduced framework connectivity at high defect concentrations must be outweighed by considerations of densification and/or local symmetry. Whatever the microscopic origin for this defect dependence, we note that the chemical control over NTE behaviour we observe here would usually be associated with aliovalent substitution, as exploited in conventional solid-state NTE compounds (\emph{e.g.}\ the ZrV$_x$P$_{2-x}$O$_7$ family\cite{Korthuis1995}). That we can achieve similar control over the physical properties of MOFs through the use of defect chemistry is precisely the proof-of-principle needed to demonstrate that defects might be used as a design element in their own right when engineering MOFs with targeted properties.

\section*{Concluding remarks}

One of the key motivations for exploring NTE behaviour in defective MOFs was always that NTE is very often diagnostic of other lattice-dynamical anomalies, such as pressure-induced amorphisation and/or mechanical softening. Consequently the variation in magnitude of NTE documented in this study implies that changes in defect concentrations are likely also to affect the mechanical stability of MOFs in high pressure environments and their resistance to amorphisation. We find that this relationship is likely to be entirely counterintuitive for defective UiOs (at the very least): reduced NTE nearly always translates to greater pressure stability, and so higher defect concentrations might now be expected actually to frustrate amorphisation rather than to encourage it. In turn this may provide a mechanism to improve the mechanical properties of MOFs, such as their machinability and durability under operating conditions. Whether this bizarre relationship of defect-driven toughening actually holds in practice remains to be shown, but what is clear even at this stage is that further studies of defect/property relationships in MOFs are set to challenge our collective intuition of the physical consequences of structural disorder.

%Difficulty of phonon calculations but would be interesting to do.---still don't say anything about this.

While we have focussed heavily on the implications of variable NTE in the UiO-66 system, we note for completeness that the control over framework densification evident in Fig.~\ref{fig4}(b) may highlight another means of exploiting defects in MOFs. It has recently been shown that densification of guest-containing MOFs through pressure or mechanical grinding can be an effective method of immobilising dangerous species (\emph{e.g.}\ radioactive iodine).\cite{Bennett2013,Sava2011} Perhaps the incorporation of defects might simultaneously increase capacity for adsorption of these guest molecules while also trapping these molecules more efficiently as a result of an increased density change on framework collapse.

%This points to the possibility that defect inclusion might facilitate encapsulation of guests etc.

In summary, we have shown for the first time how the thermomechanical properties of a MOF can be systematically varied \emph{via} controlled incorporation of defects. The thermomechanical properties we observe in defective UiO-66(Hf) are extremely unusual, and include the strongest isotropic NTE effect ever observed for a MOF. This property in itself may have application in counteracting the positive thermal expansion behaviour of other engineering materials. Yet the most important result remains a conceptual one: namely that defects might now realistically play a key role in the design of new classes of functional MOFs.

\section*{Methods}
\subsection*{Preparation of defective UiO-66(Hf)}
We synthesised defective UiO-66(Hf) using the method described for large scale synthesis in Ref.~\citenum{Cliffe2014}. The most defective sample was synthesised as follows. HfCl$_4$ (3\,mmol, 99.9\% (metals basis, $< 0.5$\% Zr), Alfa Aesar) and terephthalic acid (2.5\,mmol, Sigma Aldrich) were added to a 250\,ml Schott bottle and dissolved in N,N dimethylformamide (DMF) (40\,ml, Sigma Aldrich). This mixture was sonicated until the metal and ligand had dissolved and then formic acid (20\,ml, 95\%, Sigma Aldrich) was added. The bottle was sealed then placed in an oven at 120\,$^\circ$C for 24\,h. The product was isolated by filtration, washed (DMF) and then heated at 60\,$^\circ$C with DMF for 3 days to ensure the removal any residual ligand. Further impurities were removed from the filtered product by Soxhlet extraction (ethanol). The purified product was then dried at 150\,$^\circ$ for 24\,h \emph{in vacuo}. Less defective samples were prepared using larger concentrations of terephthalic acid: 3.0, 4.0, 4.5 and 5.0\,mmol. All samples were loaded in 0.5\,mm quartz glass capillaries with Al$_2$O$_3$ as an internal temperature standard (\emph{ca} 50\% by mass) and sealed with cotton wool.

\subsection*{Thermogravimetric Analysis}
Thermogravimetric analysis was carried using a Mettler Toledo TGA/DSC 1 System, heating from 30\,$^\circ$C to 650\,$^\circ$C at a rate of 20\,$^\circ$C\,min$^{-1}$ before cooling down to room temperature.

\subsection*{Variable-temperature synchrotron X-ray diffraction measurements}
Variable temperature X-ray powder diffraction measurements were carried out on the I11 beamline at the Diamond Light Source ($\lambda$ = 0.82562\,\AA) using a position sensitive detector with a Cyberstar hot air blower.\cite{Thompson2009,Thompson2011} Analysis of the data was carried out using Topas Academic version 4.1.\cite{Topas2007} All samples' purity were checked using a Rietveld fit to the reported structure for UiO-66(Zr),\cite{Cavka2008} with Hf substituted for Zr. The additional primitive peaks were fitted using a secondary Pawley phase with the same lattice parameter but with additional size broadening as described in Ref.~\citenum{Cliffe2014}. Lattice parameters were determined using Pawley refinement in space group $Fm \bar 3 m$, omitting the low angle region where the broad primitive reflections make an important contribution. The Pawley refinements were carried out using the previous temperature's refined parameters as the input for the next refinement.

\subsection*{Quantum Mechanical Calculations}
All structures were fully relaxed by optimizing both atomic positions and unit cell parameters. We performed quantum mechanical calculations in the density functional theory approach with localized basis sets (CRYSTAL14 code \cite{Dovesi2005}). We used the B3LYP hybrid exchange-correlation functional \cite{Becke1993} and all-electron basis sets for all atoms,\cite{Valenzano2011} except Hf for which we used a small relativistic effective core potential with 12 outer electrons considered explicitly.\cite{MunozRamo2007} Full methodological details can be found in in Refs.~\citenum{Cliffe2014} and \citenum{MunozRamo2007}. The optimised structures are included as crystallographic information files as part of the ESI\dag.

\section*{Acknowledgments}
M.J.C., J.A.H., and A.L.G. gratefully acknowledge financial support from the ERC (Grant No.~279705) and EPSRC (Grant No.~EP/G004528/2). This work was performed using HPC resources from GENCI-IDRIS (grant x2015087069). We thank Diamond Light Source for access to beamline I11 (Proposal Numbers~EE9940-1).

\balance

\footnotesize{
\bibliography{pccp_2015_uio} %your .bib file
\bibliographystyle{rsc} }

\end{document}